\documentclass[preprint,showpacs,preprintnumbers,amsmath,amssymb,aps]{revtex4}

\usepackage{graphicx}
\usepackage{dcolumn}
\usepackage{bm}
\newcommand{\vecvar}[1]{\mbox{\boldmath$#1$}}

\begin{document}

\preprint{PRESAT-8301}

\title{First-principles calculation of electronic polarization of III-V nanotubes}

\author{ Jun Otsuka, Kikuji Hirose, and Tomoya Ono}
\affiliation{Department of Precision Science and Technology, Osaka University, Suita, Osaka 565-0871, Japan}

\date{\today}

\begin{abstract}
A first-principles study of the electronic polarization of BN and AlN nanotubes and their graphitic sheets under an external electric field has been performed. We found that the polarization per atom of zigzag nanotubes increases with decreasing diameter while that of armchair nanotubes decreases. The variation of the polarization is related to the exterior angle of the bonds around the B or Al atoms rather than that around the N atoms. The increase in the polarization of the zigzag nanotubes with decreasing diameter is caused by the large variation of the exterior angle when they are wrapped into the tubular form. On the other hand, the decrease in the bond length results in the weak polarization of thin armchair nanotubes.
\end{abstract}

\maketitle

\section{Introduction}
Since the discovery of carbon nanotubes (C NTs)\cite{I0}, many experimental and theoretical studies have focused on synthesizing nanometer-scale tubular forms of various materials and revealing their properties. III-V semiconductors have also attracted considerable attention due to their many potential applications in high-power and high-temperature devices, because the properties of bulk BN are similar to those of diamond, for example, excellent thermal conductivity, high chemical resistance, and high melting point. In addition, since BN is a piezoelectric material, its tubular structure is a candidate for applications in nanometer-scale sensors and actuators. It has so far been reported that single-walled BN nanotubes (BN NTs) can be produced by arc discharge\cite{I2} and chemical substitution reactions\cite{I3}. On the theoretical side, there have been several studies on the polarization of C NTs and BN NTs\cite{Benedict1995,Novikov2006,Li2003,Brothers2005,Guo2005,Wirtz2005,Kozinsky2006,Guo2007}. For example, Kozinsky and Marzari\cite{Kozinsky2006} reported that the longitudinal polarizability of single-wall C NTs varies as the inverse square of the band gap, while the transverse polarizability is proportional to the square of the effective radius. Chen {\it et al.}\cite{I5} proved that a gradual reduction of the band gap of BN NTs occurs because of the larger ionicity of the BN bond. In addition, Guo and Lin\cite{Guo2005} and Wirtz {\it et al.}\cite{Wirtz2005} reported the liner scaling law of the longitudinal polalizabilities with respect to diameter of BN NTs. However, there still remains much to be learned about the electronic polarization of the NTs composed of III-V elements.

In this paper, we carry out a first-principles calculation on the longitudinal electronic polarization of BN and AlN NTs using the Berry phase approach\cite{4,Resta1994}. The longitudinal electronic polarization per atom of the armchair NTs decreases with decreasing diameter while the polarization of zigzag NTs increases and the polarization of the armchair NTs is smaller than that of the sheets. This tendency upon polarization cannot be explained by the conventional interpretation of the relationship between the energy band gap and the polarization\cite{2}, because the band gap of the graphitic sheets is larger than that of the NTs and the polarization is expected to increase with the reduction of the band gap. In addition, we found that the characteristics of the bond around B or Al atoms have a larger effect on the polarization than those of the bond around N atoms.

The rest of this paper is organized as follows. In Sec. II, we briefly describe the method used in this study. Our results are presented and discussed in Sec. III. We summarize our findings in Sec. IV.

\section{Computational methods}
\begin{figure}[htb]
\begin{center}
\includegraphics{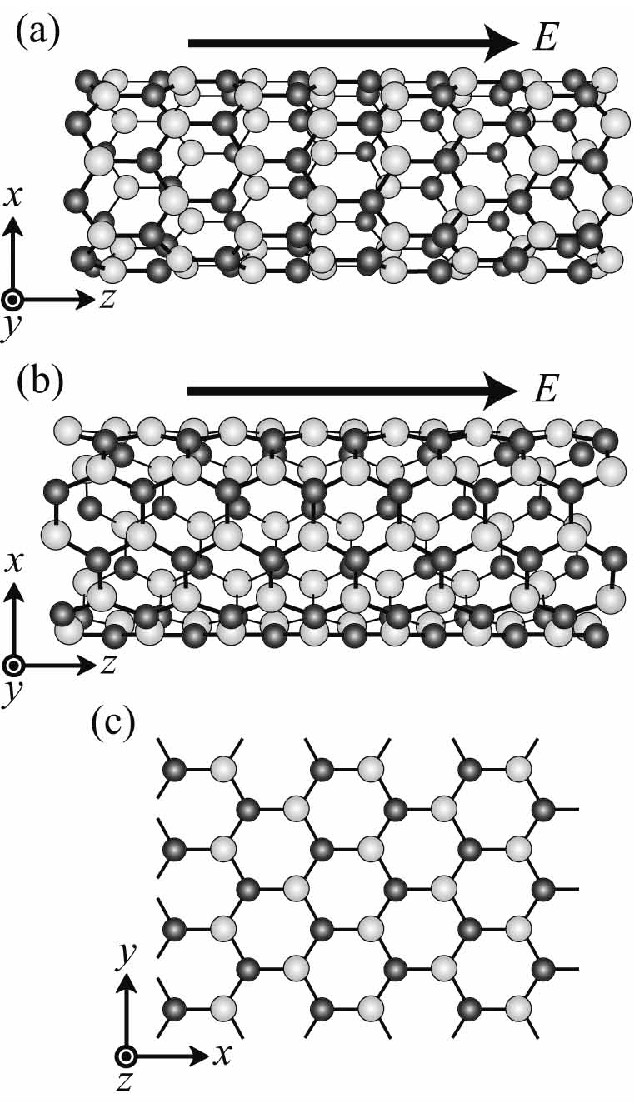}
\end{center}
\caption{Computational models of (a) zigzag NT, (b) armchair NT, and (c) graphitic sheet. Light and dark spheres indicate B and N atoms, respectively. $E$ is an external electric field.}
\label{model}
\end{figure}

Figure~\ref{model} shows the computational models of the III-V NTs and the III-V graphitic sheet. The bond length $a_0$ is set to 1.43 \AA \hspace{2mm} for BN tubes and 1.83 \AA \hspace{2mm} for AlN tubes, which are reported by the previous first-principles calculations\cite{Guo2005,Zhao}. The calculations are performed within the local density approximation\cite{Perdew1981} of density functional theory\cite{Hohenberg1964,Kohn1965} using the real-space finite-difference approach\cite{chelikowsky,Hirose2005,tsdg} and the norm-conserving pseudopotentials of Troullier and Martins in the Kleinman-Bylander representation\cite{Keinman1982,Troullier1991,kobayashi1999}. The cutoff energy is set to 110 Ry, which corresponds to a grid spacing $h$ of 0.30 a.u., and a denser grid spacing is taken to be $h/3$ in the vicinity of the nuclei with the augmentation of double-grid points\cite{tsdg,Hirose2005}. With this grid spacing, the difference in total energy from that with the grid spacing $h$ of 0.15 a.u. is 7 meV/atom. The periodic boundary condition is imposed on the supercell in all directions. During the first-principles structural optimization, we relaxed all the atoms using a conjugate gradient algorithm, reaching a tolerance in the force of $F_{max} <$ 1mH/bohr. In the case of the NTs, the length of the supercell along the tube axis is fixed at $3 a_0$ ($\sqrt{3} a_0$) for zigzag (armchair) NTs and a vacuum region of more than 10 \AA \hspace{2mm} between NTs in the neighboring supercells is taken. The integration over the Brillouin zone for the tube axis is performed by the equidistant sampling of 10 (18) {\it k}-points for the zigzag (armchair) NTs. In the case of the graphitic sheets, the supercell is $3 a_0 \times \sqrt{3} a_0 \times L_z$, where $L_z$ is the length of the supercell along the $z$ axis, which is taken to be 10 \AA \hspace{2mm} to avoid the interaction between the adjacent sheet. The sampling $\vecvar{k}$-points over the Brillouin zone is $10\times 18\times 1$. We have ensured that the enlargement of the supercell and the increase of the number of the sampling k points do not change our conclusions significantly.

An electric field of 0.1 V/\AA \hspace{2mm} is applied along the NT axis and along the directions parallel to the graphitic sheet. The electronic ground state and the optimized electronic structures under external electric fields are determined using the discrete Berry phase scheme proposed by Umari and Pasquarello\cite{4}. The electronic polarization can be expressed as
\begin{equation}
{\bf P}
=
-2e
\sum_i
\left<{\bf r}_i\right>,
\end{equation}
where $i$ is the band index, $-e$ is the electron charge, the factor $2$ indicates the spin freedom, and $\left<{\bf r}_i\right>$ is the center of the density distribution of the occupied wave functions $\psi_i$, and is defined by the following formula\cite{4}:
\begin{equation}
\left<{\bf r}_i\right>
=
-\frac{L}{2\pi}\mbox{Im}\left(\ln\left<\psi_i|e^{-i\frac{2\pi}{L}{\bf r}}|\psi_i\right>\right),
\end{equation}
with $L$ being the length of the supercell. The variation of the polarization due to the external electric field is defined by $\Delta {\bf P^E}={\bf P^E}-{\bf P^0}$, where ${\bf P^E}$ (${\bf P^0}$) corresponds to the electronic polarization when the external electric field $E$ is present (absent).

\section{Results and discussion}
\begin{figure}[htb]
\begin{center}
\includegraphics{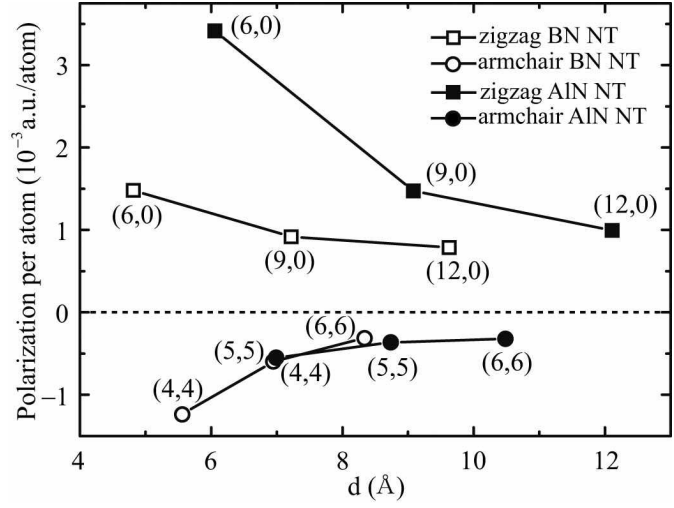}
\end{center}
\caption{Electronic polarization per atom as a function of diameter. The zero of the polarization is set to be those of the graphitic sheets, which correspond to 36.2$\times$10$^{-3}$ a.u./atom for BN NTs and 52.3$\times$10$^{-3}$ a.u./atom for AlN NTs.}
\label{graph1}
\end{figure}

Table~\ref{tbl:1} lists the electronic polarization per atom of the graphitic sheet. Although the graphitic sheets are 120-degrees rotationally symmetrical, the sheets do not exhibit in-plane anisotropy. Figure~\ref{graph1} shows the electronic polarization of the NTs. When the atomic geometries of the NTs are fully relaxed, the B or Al atoms move toward the central axis of the NTs, and the N atoms move in the opposite direction. The bond configuration around the B or Al atoms becomes planar as reported by Blase {\it et al.}\cite{blase} The radial bucklings are listed in Table \ref{tbl:2}, which are agreement with those reported by previous studies\cite{Wirtz2003,Baumeister2007}. The polarization of the zigzag NTs increases as their diameters become small. In contrast, the armchair NTs exhibit the opposite tendency upon polarization; the electronic polarization of the armchair NTs becomes smaller than that of the graphitic sheet and decreases as the NTs become thin. The different tendency of the polarization with the decrease of the diameter agrees with the results computed by Guo {\it et al.}\cite{Guo2005,Guo2007}, in which the liner scaling law of the longitudinal polalizabilities exhibits certain deviations depending on the chiralities. Although the different polarization tendency of the BN NTs was found in the previous first-principles study\cite{Guo2007}, the origin of the interesting polarization is still unsettled question. It is reported that the electronic polarization of conventional III-V compounds increases with the reduction of the band gap\cite{2}. In addition, all NTs are semiconductors and their band gap grows monotonically with increasing diameter\cite{I5}. Our results regarding the armchair NTs are inconsistent with those predicted from these previous studies because the band gap of a III-V graphitic sheet is larger than that of III-V NTs. 

\begin{figure}[htb]
\begin{center}
\includegraphics{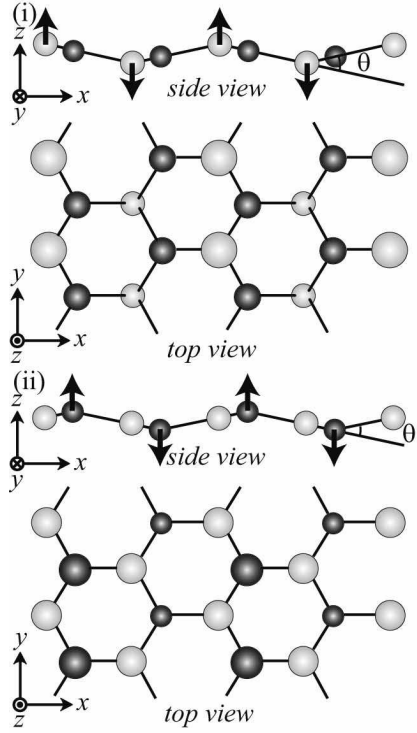}
\end{center}
\caption{Computational models of graphitic sheet in which B or N atoms are displaced along the direction perpendicular to the graphitic sheet. (i) B atoms are shifted, (ii) N atoms are shifted. Gray and dark-blue spheres indicate B and N atoms, respectively. In the top views, B(N) atoms are denoted by larger and smaller spheres according to their distance from N(B) atoms.}
\label{sheet_dz}
\end{figure}

\begin{figure}[htb]
\begin{center}
\includegraphics{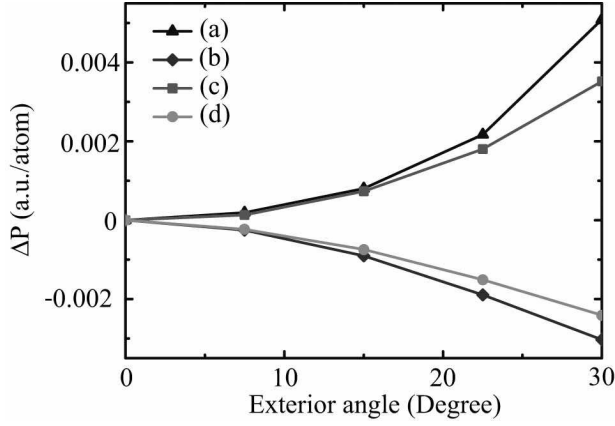}
\end{center}
\caption{Variation of electronic polarization, $\Delta {\bf P}={\bf P}(\theta)-{\bf P}(0)$, of BN graphitic sheet as a function of exterior angle $\theta$ in Fig. \ref{sheet_dz}. (a) and (b) The external electric field is applied along the $x$ direction. (c) and (d) The external electric field is applied along the $y$ direction. The curves (a) and (c)[(b) and (d)] show the polarizations of model (i)[(ii)] in Fig. \ref{sheet_dz}.}
\label{graph2}
\end{figure}

To discuss the different tendency of the polarization between the zigzag and armchair NTs in detail, we examine the polarization with respect to the BN bond angle and the cross section of the BN graphitic sheet, which varies during the wrapping of the sheet into the tubular form. We first calculate the electronic polarization of the graphitic sheet in which B or N atoms are displaced along the direction perpendicular to the graphitic sheet. Schematic illustrations of the computational models are shown in Fig.~\ref{sheet_dz}, where the neighboring B(N) atoms are alternately shifted toward the positive and negative $z$ direction. Figure~\ref{graph2} shows the polarization as a function of the exterior angle $\theta$ of the BN bonds. The polarization increases (decreases) nearly quadratically with the exterior angle around the B(N) atoms, and the variation of the exterior angle around the B atoms has a stronger effect on the increase in the polarization than that around the N atoms. Thus, the polarization becomes totally stronger as the exterior angle increases.

\begin{figure}[htb]
\begin{center}
\includegraphics{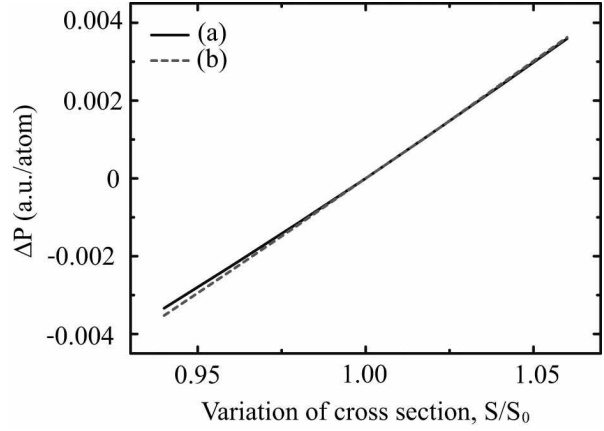}
\end{center}
\caption{Variation of electronic polarization, $\Delta {\bf P}={\bf P}(S)-{\bf P}(S_0)$, of BN graphitic sheet as a function of variation of cross section. The computational model is the same as that in Fig. \ref{model} (c). The curve (a)[(b)] shows the polarization of the case that the external electric field is applied along the $x[y]$ direction and the length of the supercell in the $y[x]$ direction is varied.}
\label{graph3}
\end{figure}

We next explore the polarization properties with respect to the BN bond length. The electronic polarization of the graphitic sheet as a function of the ratio of the cross section of the sheets, $S/S_0$, is shown in Fig.~\ref{graph3}. Here, $S=tL_{x(y)}$, $t$ is the effective thickness of the sheet\cite{comment}, and $L_{x(y)}$ is the length of the supercell in the $x(y)$ direction. In addition, $S_0$ is the cross section of the sheet in which the lengths of all the BN bonds are $a_0$. The length of the supercell in the direction perpendicular to the external electric field is varied, while that in the direction parallel to the field is fixed and the graphitic sheet is kept flat. For both directions of the external electric field, the polarization proportionally becomes large with increasing lateral length $L_{x(y)}$.

\begin{figure}[htb]
\begin{center}
\includegraphics{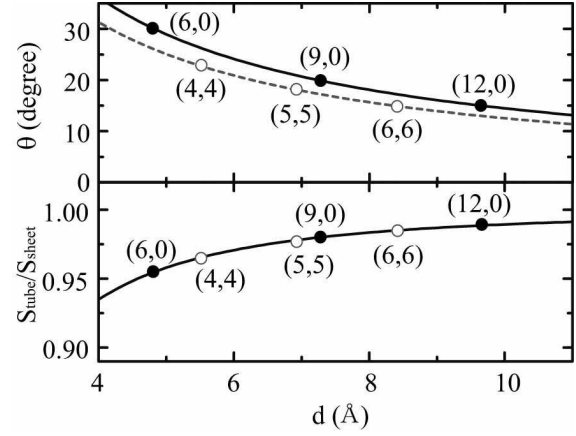}
\end{center}
\caption{Variations of exterior angle of the BN bonds and cross section of the BN NTs as a function of diameter of NTs. The solid and dotted curves are the variations of the zigzag and armchair NTs, respectively.}
\label{graphx}
\end{figure}

To explore the relation between the polarizability and the geometrical deformation due to the wrapping into the tubular form, we plot in Fig.~\ref{graphx} the variations of the exterior angle of the BN bonds and the cross section of the BN NTs as a function of the diameter. When the BN graphitic sheet is wrapped into the tubular form, the exterior angle increases and the cross section becomes small with the decrease in the diameter. The external angle of the zigzag NTs varies more than that of the armchair NTs; the large variation of the exterior angle increases the polarization of the zigzag NTs. In contrast, the decrease in the cross section results in the weak polarization of thin armchair NTs.

\section{Conclusion}
We have studied the electronic polarization of BN and AlN NTs and their graphitic sheets under finite external electric fields from first principles. We found that the electronic polarization per atom of the zigzag NTs increases as their diameter decreases. On the contrary, the polarization of the armchair NTs decreases as the NTs become thin. The variation of the polarization with the diameter is associated with specific features of the bond arrangement, such as the bond angles around the B or Al atoms and the bond length. Our results indicate the possibility of tuning the dielectric property of III-V NTs by controlling their diameter and chirality.

\section*{Acknowledgements}
This research was partially supported by a Grant-in-Aid for the 21st Century COE ``Center for Atomistic Fabrication Technology'', by a Grant-in-Aid for Scientific Research in Priority Areas ``Development of New Quantum Simulators and Quantum Design'' (Grant No. 17064012), and also by a Grant-in-Aid for Young Scientists (B) (Grant No. 17710074) from the Ministry of Education, Culture, Sports, Science and Technology. The numerical calculation was carried out using the computer facilities at the Institute for Solid State Physics at the University of Tokyo, the Research Center for Computational Science at the National Institute of Natural Science, and the Information Synergy Center at Tohoku University.

\begin{table}[htbp]
\caption{Polarization per atom (a.u./atom) of graphitic sheets.}
\label{tbl:1}
\begin{center}
\begin{tabular}{c|cc}
\hline\hline 
 & \hspace{3mm}BN\hspace{3mm} & \hspace{3mm}AlN\hspace{3mm} \\ \hline
Polarization per atom & 0.0362 & 0.0523 \\ \hline\hline
\end{tabular}
\end{center}
\end{table}

\begin{table}[htbp]
\caption{Radial bucklings (\AA) of III-V nanotubes.}
\label{tbl:2}
\begin{center}
\begin{tabular}{c|cc}
\hline\hline 
 & \hspace{3mm}BN\hspace{3mm} & \hspace{3mm}AlN\hspace{3mm} \\ \hline
(6,0) & 0.090 & 0.119 \\
(9,0) & 0.057 & 0.083 \\
(12,0) & 0.042 & 0.062 \\
(4,4) & 0.080 & 0.113 \\
(5,5) & 0.061 & 0.091 \\
(6,6) & 0.050 & 0.074 \\ \hline\hline
\end{tabular}
\end{center}
\end{table}


\begin{thebibliography}{99}
\bibitem{comment}
The effective thickness $t$ can be taken to be arbitrary because it is canceled in $(S-S_0)/ S_0$.
\end{thebibliography}
\end{document}